\documentclass[conference]{IEEEtran}
\IEEEoverridecommandlockouts
\usepackage{cite}
\usepackage{amsmath,amssymb,amsfonts}
\usepackage{algorithmic}
\usepackage{graphicx}
\usepackage{textcomp}
\usepackage{xcolor}
\usepackage[dvipsnames]{xcolor}
\def\BibTeX{{\rm B\kern-.05em{\sc i\kern-.025em b}\kern-.08em
    T\kern-.1667em\lower.7ex\hbox{E}\kern-.125emX}}

\usepackage{pgfplots}
\usepackage{subcaption}  
\usepackage{float}
\usepackage{tikz}
\usepackage{pgfplotstable}
\usepackage{changepage}
\usepackage{caption}
\usepackage{bm}
\usepgfplotslibrary{groupplots}
\usepackage{stfloats}
\usepackage{enumitem}

\usepackage[shortcuts, acronym, toc, nonumberlist=false]{glossaries}
\makenoidxglossaries
\newacronym{csi}{CSI}{channel state information}
\newacronym{dl}{DL}{downlink}
\newacronym{em}{EM}{expectation–maximization}
\newacronym{foh}{FOH}{first-order hold}
\newacronym{gmm}{GMM}{Gaussian mixture model}
\newacronym{gru}{GRU}{gated recurrent unit}
\newacronym{lisa}{LISA}{linear successive allocation}
\newacronym{lmmse}{LMMSE}{linear minimum mean squared error}
\newacronym{los}{LOS}{line-of-sight}
\newacronym{mi}{MI}{mutual information}
\newacronym{mimo}{MIMO}{multiple-input multiple-output}
\newacronym{ml}{ML}{machine learning}
\newacronym{mse}{MSE}{mean squared error}
\newacronym{mu}{MU}{multi-user}
\newacronym{nmse}{NMSE}{normalized mean squared error}
\newacronym{nn}{NN}{neural network}
\newacronym{ofdm}{OFDM}{orthogonal frequency-division multiplexing}
\newacronym{sinr}{SINR}{signal-to-interference-plus-noise ratio}
\newacronym{snr}{SNR}{signal-to-noise ratio}
\newacronym{svd}{SVD}{singular value decomposition}
\newacronym{tgn}{TGn}{Task Group n}
\newacronym{wifi}{Wi-Fi}{wireless fidelity}
\newacronym{zoh}{ZOH}{zero-order hold}

\makeatletter
\renewcommand\section{\@startsection{section}{1}{0pt}%
  {1.2ex}
  {0.8ex}
  {\bfseries\centering}}
\renewcommand\subsection{\@startsection{subsection}{2}{0pt}%
  {1ex}
  {0.5ex}
  {\bfseries}}
\makeatother

\usepackage{fancyhdr}
\fancypagestyle{cfooter}{
  \fancyhf{}

  \cfoot{\scriptsize This work has been submitted to the IEEE for possible publication. 
  Copyright may be transferred without notice, after which this version may no longer be accessible.}
}

\begin{document}
\pagestyle{cfooter}

\title{Efficient Channel Prediction based on Gram-Square-Root Factorization using GMMs
\thanks{This work was funded by Toga Networks Ltd., Hod HaSharon, Israel.}}

\author{Kathrin Klein, Amar Kasibovic, Michael Joham, Shachar Shayovitz$^\dagger$, Wolfgang Utschick\\
\IEEEauthorblockA{School of Computation, Information and Technology, Technical University of Munich, Germany\\
$^\dagger$Toga Networks Ltd., Hod HaSharon, Israel\\
Email: \{kathrin.klein; amar.kasibovic; joham; utschick\}@tum.de\\
	$^\dagger$Email: shachar.shayovitz@huawei.com}}

\maketitle

\setlength{\abovedisplayskip}{4pt}
\setlength{\belowdisplayskip}{4pt}
\setlength{\abovedisplayshortskip}{2pt}
\setlength{\belowdisplayshortskip}{2pt}

\begin{abstract}
Accurate \gls{csi} is critical for \gls{dl}-\gls{mu}-\gls{mimo} systems, where feedback delays and mobility can degrade precoding performance.
To ensure reliable beamforming and interference mitigation, \gls{csi} prediction is required.
In practical systems, full \gls{csi} feedback is often infeasible due to signaling overhead, so transmitters rely on partial \gls{csi} reported by the receivers.
In this work, we propose a \gls{gmm}-based prediction framework for \gls{mimo}-\gls{ofdm} channels under partial feedback using Gram-square-root factorization.
To address the high dimensionality, we introduce an efficient parameter reduction technique that exploits structured covariance matrices, significantly lowering complexity without noticeable performance degradation.
This reduction is based on the Gram-square-root factorization and remains of interest even when full \gls{csi} is available.
Simulation results demonstrate that \glspl{gmm} achieve the highest prediction accuracy and correctly capture the underlying channel subspaces, which is essential for effective \gls{mu}-precoding.
The proposed method outperforms classical baselines such as \gls{zoh}, \gls{foh}, and \gls{lmmse} predictors, and an advanced \gls{nn}-based predictor.
Notably, the parameter-reduced partial \gls{csi} \gls{gmm} achieves performance comparable to that of full \gls{csi} prediction, highlighting its ability to efficiently model the channel structure under limited feedback.
\end{abstract}


\section{Introduction and Motivation}
\Gls{dl}-\gls{mu}-\gls{mimo} transmission is a key enabler for high throughput in modern wireless communication systems.
By allowing a transmitter to serve multiple users simultaneously, these systems significantly improve spectral efficiency and overall network capacity.
Precoding strategies, which spatially separate user signals and minimize inter-user interference, rely on accurate channel knowledge at the transmitter like in \cite[Sec.~13.7]{ieee-802-11n, ieee-802-11n-book}.
In indoor scenarios, even under low mobility conditions with user speeds on the order of 1--3~km/h, the delay between channel estimation and the actual data transmission introduces a mismatch between the estimated and the true \gls{csi}.
This delay, often on the order of several milliseconds due to channel sounding, feedback, and processing, allows the wireless channel to evolve because of user motion or environmental changes.
As a result, the previously acquired \gls{csi} may become outdated, leading to reduced beamforming gains, increased inter-user interference, and overall performance degradation.
To address this, robust prediction of future \gls{csi} is required.
Traditional approaches, such as \gls{zoh} or \gls{foh} (also called linear extrapolation) \cite[Sec.~2]{oppenheim2010discrete}, offer low complexity but rapidly degrade in the presence of mobility and for longer prediction horizons.
Other methods like prediction using \gls{lmmse} \cite[Sec.~9.1.6]{hossein2014nmselslmmsebook} exploit statistical correlations between past and future \gls{csi}, but they rely on Gaussian assumptions that often do not hold in real-world channels.

In practical systems, an additional challenge arises in the \gls{dl}, where the transmitter relies on channel information estimated by the receivers and only a compressed representation of the \gls{csi} is reported leading to partial \gls{csi} at the transmitter.
A prominent example of such a system is \gls{wifi}, where IEEE~802.11 standards \cite{ieee-802-11n} rely on compressed feedback mechanisms to limit signaling overhead.
In these systems, each receiver estimates its \gls{dl} channel and feeds back a compressed representation containing dominant singular values and the corresponding right singular vectors, while the left singular vectors are not reported.
Consequently, the transmitter reconstructs only an approximate, limited representation of the channel.
This introduces a broader challenge that arises in many practical feedback limited \gls{mimo} systems.

Previous work on channel prediction using \gls{ml} has largely focused on cellular networks as in \cite{kim2025machinelearningfuturewireless, turan2024gmmchannelprediction}, often providing forward-looking discussions on \gls{mimo}-\gls{ofdm} scenarios without concrete implementation.  
Importantly, these approaches typically assume access to full \gls{csi}, and thus do not account for the partial feedback.
In this work, we investigate \glspl{gmm} for partial \gls{csi} prediction.
Our main contributions are summarized as follows:
\begin{itemize}
    \item We propose a prediction framework for forecasting future \gls{csi} in wireless, high-dimensional \gls{dl}-\gls{mu}-\gls{mimo}-\gls{ofdm} systems.
    \item Channel prediction is performed under partial, limited feedback using Gram-square-root factorization.
    \item We introduce a parameter reduction technique that leverages structural properties of the covariance matrix in the presence of partial feedback, without significant performance loss.
    \item We demonstrate that the proposed approach outperforms conventional channel prediction baselines including \glspl{nn}.
\end{itemize}

\section{System Model}\label{sec:system-model}
Let the tensor $\bm{H} \in \mathbb{C}^{N_{\text{t}} \times N_{\text{f}} \times N_{\text{rx}} \times N_{\text{tx}}}$ be a \gls{mimo}-\gls{ofdm} channel, where $N_{\text{tx}}$ and $N_{\text{rx}}$ denote the number of transmit and receive antennas, respectively.
$N_{\text{f}}$ is the number of subcarriers and $N_{\text{t}}$ the number of time steps.
For channel prediction, we consider that each data sample consists of $N_{\text{t}} = N_{\text{t,o}} + N_{\text{t,p}}$ time steps, where $N_{\text{t,o}}$ represents the number of observed (past) time steps and $N_{\text{t,p}}$ the number of predicted (future) time steps.
The dataset of vectorized channels $\bm{h}^{(\ell)} = \text{vec}(\bm{H}^{(\ell)}) \in \mathbb{C}^{N_{\text{total}}}$ is denoted as $\mathcal{H} = \{ \bm{h}^{(\ell)} \}_{\ell=1}^{L}$, where $L$ is the dataset's total number of channel realizations (samples) and $N_{\text{total}} = N_{\text{t}} N_{\text{f}} N_{\text{rx}} N_{\text{tx}}$ denotes the total dimension of each data sample.
Each sample $\bm{h}^{(\ell)}$ can be decomposed into the observed channel $\bm{h}_{\text{o}}^{(\ell)}$ and the predicted channel $\bm{h}_{\text{p}}^{(\ell)}$:
\begin{equation}
    \bm{h}^{(\ell)} =
    \begin{bmatrix} 
        \bm{h}_{\text{o}}^{(\ell)} \\ 
        \bm{h}_{\text{p}}^{(\ell)} 
    \end{bmatrix},
    \quad 
    \bm{h}_{\text{o}}^{(\ell)} \in \mathbb{C}^{N_{\text{total,o}}}, \quad
    \bm{h}_{\text{p}}^{(\ell)} \in \mathbb{C}^{N_{\text{total,p}} },
\end{equation}
where $N_{\text{total,o}} = N_{\text{t,o}} N_{\text{f}} N_{\text{rx}} N_{\text{tx}}$ and $N_{\text{total,p}} = N_{\text{t,p}} N_{\text{f}} N_{\text{rx}} N_{\text{tx}}$.
We model noisy channel observations as $\bm{y}_{\text{o}} = \bm{h}_{\text{o}} + \bm{n}$, where $\bm{n} \sim \mathcal{N}_{\mathbb{C}}(\bm{0}, \sigma^2 \bm{I})$ represents additive Gaussian noise.
In the following, we use the notation $\hat{\bm{h}}_{\text{p}}$ for the prediction, which estimates the true future channel $\bm{h}_{\text{p}}$ based on the noisy channel observation $\bm{y}_{\text{o}}$.

\section{Preliminaries on Channel Prediction using GMMs}
Based on the system model defined in Section~\ref{sec:system-model}, we briefly summarize the approach in \cite{turan2024gmmchannelprediction}, which serves as the foundation for our model.
Given training data at arbitrary \glspl{snr}, we can train a \gls{gmm} offline and later use it online to predict future channels from past noisy observations.

\subsection{Offline Training}\label{sec:gmm-offline-traning}
A \gls{gmm} models the distribution of a wireless channel $\bm{h}$ as a mixture of $K$ Gaussian components \cite[Sec.~9.2-9.4]{bishop2006}, each with a mixture weight $\pi_k$, a mean $\bm{\mu}_k$, and a covariance $\bm{C}_k$:
\begin{equation}
    p(\bm{h}) = \sum_{k=1}^K \pi_k \mathcal{N}_{\mathbb{C}}(\bm{h}; \bm{\mu}_k, \bm{C}_k),
\end{equation}
with latent variable $k$ indicating the component.
Model parameters are optimized using the \gls{em} algorithm, iterating between an E-step and a M-step.
At iteration $i$, the E-step calculates a so-called "responsibility" of component $k$, for a given sample $\bm{h}$ \cite[Sec.~9.2-9.4]{bishop2006} as
\begin{equation}
    p_i(k|\bm{h}) = \frac{\pi_k \mathcal{N}_{\mathbb{C}}(\bm{h}; \bm{\mu}_k, \bm{C}_k)}{\sum_{j=1}^K \pi_j \mathcal{N}_{\mathbb{C}}(\bm{h}; \bm{\mu}_j, \bm{C}_j)}.
\end{equation}
Given the responsibilities, in the M-step, the parameters  $\{\pi_k, \bm{\mu}_k, \bm{C}_k\}$ for each component $k$ are updated to maximize the expected complete-data log-likelihood as described in \cite[Sec.~9.2-9.4]{bishop2006}.
The algorithm alternates between E- and M-step until convergence of the log-likelihood.
After convergence, the $k$-th component encodes the following temporal structure:
\begin{equation}
\bm{\mu}_k =
\begin{bmatrix}
\bm{\mu}_{k,\bm{h}_\text{o}} \\ \bm{\mu}_{k,\bm{h}_\text{p}}
\end{bmatrix}, \quad
\bm{C}_k =
\begin{bmatrix}
\bm{C}_{k,\bm{h}_\text{o}\bm{h}_\text{o}} & \bm{C}_{k,\bm{h}_\text{o}\bm{h}_\text{p}} \\ \bm{C}_{k,\bm{h}_\text{p}\bm{h}_\text{o}} & \bm{C}_{k,\bm{h}_\text{p}\bm{h}_\text{p}}
\end{bmatrix},
\end{equation}
where the subscript denotes the component index $k$ and the associated random variables, here $\bm{h}_\text{o}$ and $\bm{h}_\text{p}$.

\subsection{Online Channel Prediction}
Given an observation $\bm{y}_{\text{o}}$, the prediction of the future channel $\hat{\bm{h}}_{\text{p}}$ is obtained via the conditional expectation \cite{turan2024gmmchannelprediction}
\begin{equation}\label{eq:gmm-prediction}
\hat{\bm{h}}_{\text{p,GMM}} = \mathbb{E}[\bm{h}_{\text{p}} \mid \bm{y}_{\text{o}}] = \sum_{k=1}^K p(k \mid \bm{y}_{\text{o}})\, \mathbb{E}[\bm{h}_{\text{p}} \mid \bm{y}_{\text{o}}, k],
\end{equation}
with posterior weights and conditional means
\begin{equation}
\begin{aligned}
    &p(k \mid \bm{y}_{\text{o}}) = \frac{\pi_k \, \mathcal{N}_{\mathbb{C}}(\bm{y}_{\text{o}}; \bm{\mu}_{k,\bm{h}_\text{o}}, \bm{C}_{k,\bm{h}_\text{o}\bm{h}_\text{o}} + \sigma^2 \bm{I})}{\sum_{j=1}^K \pi_j \, \mathcal{N}_{\mathbb{C}}(\bm{y}_{\text{o}}; \bm{\mu}_{j,\bm{h}_\text{o}}, \bm{C}_{j,\bm{h}_\text{o}\bm{h}_\text{o}} + \sigma^2 \bm{I})}, \quad \\
    &\mathbb{E}[\bm{h}_{\text{p}} \mid \bm{y}_{\text{o}}, k] = \\
    & \qquad \qquad \bm{C}_{k,\bm{h}_\text{p}\bm{h}_\text{o}} (\bm{C}_{k,\bm{h}_\text{o}\bm{h}_\text{o}} + \sigma^2 \bm{I})^{-1} (\bm{y}_{\text{o}} - \bm{\mu}_{k,\bm{h}_\text{o}}) + \bm{\mu}_{k,\bm{h}_\text{p}}.
\end{aligned}
\end{equation}
Using~\eqref{eq:gmm-prediction} allows robust, noise-aware prediction, capturing non-linear dependencies between past and future channels.

\section{Partial Channel Prediction using GMMs}
In many standardized \gls{dl} systems that employ spatial precoding, the transmitter relies on receivers’ feedback containing only partial information about the spatial \gls{csi}.
This is the case, for example, in IEEE 802.11n \cite{ieee-802-11n} and later standards.
In particular, the receiver estimates the full spatial \gls{csi} given by $\bm{H}_{t,f} \in \mathbb{C}^{N_{\text{rx}} \times N_{\text{tx}}}$ for each time step $t$ and subcarrier $f$, and compresses it using a \gls{svd}.
Only a reduced approximation based on the dominant modes is fed back, where the number of retained dominant modes is $N_{\text{s}} \le \min(N_{\text{rx}}, N_{\text{tx}})$.
The feedback consists of the diagonal matrix $\bm{S}_{t,f} = \mathrm{diag}(s_{t,f,1}, \ldots, s_{t,f,N_{\text{s}}}) \in \mathbb{C}^{N_{\text{s}} \times N_{\text{s}}}$, where $s_{t,f,i}$ denote the $i$-th singular value, along with the corresponding right singular vectors $\bm{V}_{t,f} \in \mathbb{C}^{N_{\text{tx}} \times N_{\text{s}}}$.
The corresponding left singular vectors $\bm{U}_{t,f} \in \mathbb{C}^{N_{\text{rx}} \times N_{\text{s}}}$, which describe the receive-side spatial directions, are not transmitted, leaving the transmitter with a partial view of the \gls{dl} channel for precoding \cite[Sec.~13.7]{ieee-802-11n, ieee-802-11n-book}.
Therefore, we must operate on partial \gls{csi}, making traditional full \gls{csi} approaches unsuitable and motivating a \gls{ml} approach to predict the future $\bm{S}_{t,f}$ and $\bm{V}_{t,f}$.

In the following, we consider only partial channel feedback.
Therefore, we define the partial \gls{csi} for a specific time step $t$ and subcarrier $f$ as:
\begin{equation}\label{eq:partail-csi-def}
\bm{Z}_{t,f} = \bm{S}_{t,f} \bm{V}_{t,f}^{\operatorname{H}} \in \mathbb{C}^{N_{\text{s}} \times N_{\text{tx}}}.
\end{equation}
Here, $\bm{Z}_{t,f}$ can be interpreted as the square-root of the (reduced) Gram matrix of the full channel, because $\bm{H}_{t,f}^{\operatorname{H}}\bm{H}_{t,f} = \bm{Z}_{t,f}^{\operatorname{H}}\bm{Z}_{t,f}$ holds.
Each $\bm{Z}_{t,f}$ is vectorized along spatial dimensions and subcarriers resulting in $\bm{z}_t \in \mathbb{C}^{N_{\text{f}} N_{\text{s}} N_{\text{tx}}}$.
This now gives the new data sample $\bm{z}_t$, from which we can obtain the partial channel observation vector for $N_{\text{t,o}}$ past time steps $\bm{z}_{\text{o}} $ and the prediction target for $N_{\text{t,p}}$ future time steps $\bm{z}_{\text{p}}$, as performed in Section~\ref{sec:system-model} for the full \gls{csi}.
We can then train a \gls{gmm} as described in Section~\ref{sec:gmm-offline-traning} using $\bm{z}$.
For the online prediction phase we can infer $\hat{\bm{z}}_{\text{p}}$ from a noisy partial channel observation $\tilde{\bm{y}}_{\text{o}} = \bm{z}_{\text{o}} + \bm{n}$ as
\begin{equation}\label{eq:partial-channel-prediction}
\begin{aligned}
\hat{\bm{z}}_{\text{p,GMM}} &= \mathbb{E}[\bm{z}_{\text{p}} \mid \tilde{\bm{y}}_{\text{o}}]
=\sum_{k=1}^K p(k \mid \tilde{\bm{y}}_{\text{o}})\,\\
&\bm{C}_{k,\bm{z}_{\text{p}}\bm{z}_{\text{o}}} (\bm{C}_{k,\bm{z}_{\text{o}}\bm{z}_{\text{o}}} + \sigma^2 \bm{I})^{-1} (\tilde{\bm{y}}_{\text{o}} - \bm{\mu}_{k,\bm{z}_{\text{o}}}) + \bm{\mu}_{k,\bm{z}_{\text{p}}},
\end{aligned}
\end{equation}
where the posterior weights are determined as
\begin{equation}
    p(k \mid \tilde{\bm{y}}_{\text{o}}) = \frac{\pi_k \, \mathcal{N}_{\mathbb{C}}(\tilde{\bm{y}}_{\text{o}}; \bm{\mu}_{k,\bm{z}_\text{o}}, \bm{C}_{k,\bm{z}_\text{o}\bm{z}_\text{o}} + \sigma^2 \bm{I})}{\sum_{j=1}^K \pi_j \, \mathcal{N}_{\mathbb{C}}(\tilde{\bm{y}}_{\text{o}}; \bm{\mu}_{j,\bm{z}_\text{o}}, \bm{C}_{j,\bm{z}_\text{o}\bm{z}_\text{o}} + \sigma^2 \bm{I})}.
\end{equation}
Here, a joint representation of the singular values and right singular vectors was chosen to save parameters.
Furthermore, $\bm{Z}_{t,f}$ possesses properties that allow for a further reduction of the parameters to be learned, as explained in the following.

\subsection{Complexity Reduction}\label{sec:block-diag-t}
In the case of full \gls{csi}, each \gls{gmm} component's covariance consists of $N_{\text{total}}^{2}$ real-valued parameters.
Such number rapidly scales for large \gls{mimo}-\gls{ofdm} systems.
To reduce complexity, we can exploit structural properties of wireless channels.
The uniform separation of timestamps, the uniform subcarrier spacing and uniform antenna spacing at the transmitter and receiver, result in Toeplitz-structured covariance matrices in each of these domains individually, and an overall nested block-in-block Toeplitz covariance.
The resulting nested 4D block-in-block Toeplitz matrix drastically reduces parameters.
As an example, for a setup using $N_\text{t}=5$, $N_\text{f}=16$, $N_\text{rx}=2$ and $N_\text{tx}=8$, we reduce from $1{,}638{,}400$ to $20{,}480$ parameters ($80\times$ reduction), see Table~\ref{tab:partial-csi-covariance-compare}, where the count of the parameters is performed using \cite{fesl2022gmmcovstructures}.
In case of partial \gls{csi}, time and frequency covariances remain Toeplitz-structured, while we empirically observe that the spatial covariance $\bm{Z}_{t,f}$, for a specific time step $t$ and subcarrier $f$, exhibits a block-diagonal pattern with (near-)Toeplitz blocks, where the off diagonals were observed to be zero matrices.
One possible explanation follows from favorable propagation effects, where increasing the number of antennas (e.g., from 2 to 8) makes the effect more pronounced.
In this context, $\bm{U}_{t,f}$ and $\bm{V}_{t,f}$ inherit structural properties from the receiver and transmitter covariance matrices, respectively, and can therefore be assumed to follow their statistics, leading to zero means and Toeplitz-structured covariances on both sides.
Furthermore, the singular vectors are uncorrelated and thus the covariance of the vectorized~\eqref{eq:partail-csi-def} approximates to
\begin{equation}
\mathrm{Cov}[\mathrm{vec}(\bm{Z}_{t,f})] \approx 
\mathrm{diag}(\tilde{\bm{C}}_{t,f,1},\ldots,\tilde{\bm{C}}_{t,f,N_{\text{s}}}),
\end{equation}
where each block is defined as $\tilde{\bm{C}}_{t,f,i} = \mathbb{E}\!\left[s_{t,f,i}^2 \, \bm{v}_{t,f,i}\,\bm{v}_{t,f,i}^{\operatorname{H}}\right]$, where $s_{t,f,i}$ denotes the $i$-th singular value and $\bm{v}_{t,f,i}$ the corresponding right singular vector.
For the previously considered example, this block-diagonal Toeplitz model requires only $10{,}240$ parameters per component, a $160\times$ reduction.
This is an even stronger reduction than the 4D block-in-block Toeplitz structure as shown in Table~\ref{tab:partial-csi-covariance-compare}.
This implies that using partial \gls{csi} and exploiting the block-diagonal Toeplitz covariance structure is also interesting for systems where full \gls{csi} is available, due to the parameter reduction.

To keep the model as compact as possible and enable training with limited data while avoiding overfitting, we further reduce the dimensionality by treating each \gls{ofdm} subcarrier as an independent, time-varying narrowband \gls{mimo} channel realization.
For frequency-selective channels, a full cross-subcarrier covariance matrix would require $N_\text{f}^2$ parameters.
Instead, feeding each subcarrier vector $\bm{z}_{f} \in \mathbb{C}^{N_\text{t} N_\text{rx} N_\text{tx}}$ separately as an independent channel realization into the \gls{gmm} omits cross-subcarrier terms.
The per-component covariance parameter count becomes $4K N_\text{t}  N_\text{rx} N_\text{tx}$.
Using the example dimensions mentioned in Section~\ref{sec:block-diag-t}, this calculates to $320$ covariance parameters for the considered setup when combined with block-diagonal Toeplitz covariance-structure modeling, resulting in a $5{,}120\times$ reduction.

\begin{table}[!t]
\centering
\begin{tabular}{|c|c|c|c|}
\hline
\textbf{Covariance Type} & \textbf{Parameter Count Covs.} & \textbf{Example}\\
\hline
Full & $K N_\text{t}^2 N_\text{f}^2 N_\text{rx}^2 N_\text{tx}^2$ & 1,638,400 \\
4D-Block-Toeplitz & $16K N_\text{t} N_\text{f} N_\text{rx} N_\text{tx}$ & 20,480 \\
Block-Diag.-Toeplitz & $8K N_\text{t} N_\text{f} N_\text{rx} N_\text{tx}$ & 10,240 \\
\hline
\end{tabular}
\caption{Covariance Matrix Parameter Count Comparison for $K=1$, $N_\text{t}=5$, $N_\text{f}=16$, $N_\text{rx}=2$ and $N_\text{tx}=8$.}
\label{tab:partial-csi-covariance-compare}
\vspace{-0.5cm}
\end{table}

\section{Simulation Results on TGn Model D}
For evaluating the proposed method, we consider the \gls{tgn} family of models \cite{erceg2004tgn}, which model an indoor clustered multi-path propagation environment for \gls{wifi} systems.
Standard implementations of the \gls{tgn} channel model D, as \cite{matlabwlan2025}, define a two-dimensional propagation environment with fixed cluster center angles, a single environmental velocity, and static \gls{los} angles.
According to \cite{boeck2025wirelesschannelmodelingforml}, these properties lead to approximately Gaussian data, an assumption that represents a strong simplification of real wireless propagation environments and makes the use of \gls{ml} models unnecessary.
To address this, we modify the channel model by introducing stochastic variations, thereby better capturing realistic channel behavior.
Specifically, for each propagation cluster, the cluster center azimuth departure and arrival angles are sampled from $\mathcal{U}(0^\circ, 360^\circ)$, and the environmental velocity is drawn from $\mathcal{U}(0, 2.4{\text{km/h}})$, while cluster and ray arrival rates are modeled as Poisson process with the power following a double exponential decay law, following the guidelines from \cite{erceg2004tgn}.
Similarly, a \gls{los} is added considering a random phase shift, zero delay, with random azimuth angles and environmental velocity sampled as for the propagation clusters.
These modifications yield channel realizations that build a more realistic overall distribution.

The evaluated system uses $N_{\text{tx}}=8$ transmit and $N_{\text{rx}}=2$ receive antennas, with $N_{\text{f}}=16$ subcarriers and sequences of $N_{\text{t}}=5$ time steps for prediction.
The carrier frequency is set to 5.16~GHz, and the subcarrier spacing is 312.5~kHz.
We will discuss the setting $(N_{\text{t,o}},N_{\text{t,p}})=(2,3)$, with symbol durations $\Delta t\in\{1,10,20\}$\,ms.
Each dataset contains $100{,}000$ training and $10{,}000$ test samples.
Performance is analyzed over different \glspl{snr} for various numbers of Gaussian components up to $K=256$ and prediction steps.
The prediction steps, defined for $t > N_{\text{t,o}}$, are denoted as $t_{\text{p}} = t - N_{\text{t,o}}$.

\begin{figure*}[!t] 
    \centering
    \ref{sharedLegend}
    \vspace{-0.4cm}
\end{figure*}
\begin{figure*}[!t]
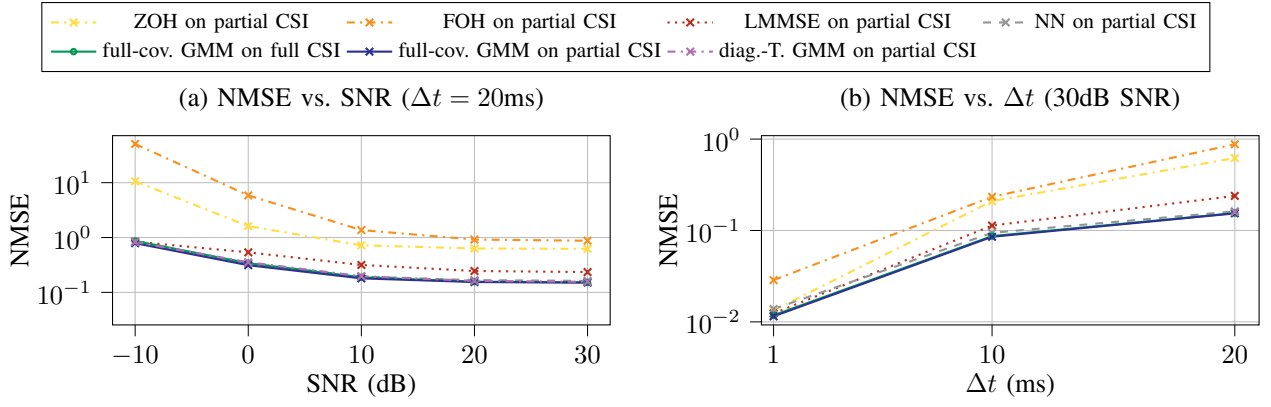
 
    \centering
    \begin{tikzpicture}
        \begin{groupplot}[
            group style={
                group size=2 by 1,
                horizontal sep=2cm,
            },
            width=0.45\textwidth,
            height=0.225\textwidth,
            legend to name=sharedLegend,
            legend columns=4,
            legend style={fill opacity=0.8, draw opacity=1, text opacity=1, anchor=south, font=\footnotesize}
        ] 
        \input{images/tgn_nmse_vs_snr}
        \input{images/tgn_nmse_vs_dt}
        \end{groupplot}
    \end{tikzpicture}
    \caption{Evaluation of \gls{nmse} over SNR and over $\Delta t$: $N_{\text{t,o}}=2$, $N_{\text{t,p}}=3$, pred. step $t_{\text{p}}=1$ and $K=256$ for the \gls{gmm} variants.}
    \label{fig:tgn-nmse}
    \vspace{-0.6cm}
\end{figure*}

\subsection{Evaluation Metrics}
We evaluate prediction performance using the \gls{nmse} as defined in \cite[Sec.~9.1.5]{hossein2014nmselslmmsebook}, which assesses reconstruction accuracy via the Frobenius-norm distance between predicted $\hat{\bm{z}}_{\text{p},t}$ and true partial \gls{csi} $\bm{z}_{\text{p},t}$ for all test samples at a time step $t$.
For the \gls{mu} setup, we consider the sum-rate metric, as in \cite[Sec.~9]{cover2006inftheory}.
We use the predicted partial \gls{csi} $\bm{Z}_{t,f}^{(\ell)}$ to determine the precoder for each test sample $\ell$, time step $t$ and subcarrier $f$.
We evaluate the results across varying numbers of users and considered different precoding strategies, including \gls{lisa} precoding, which maximizes the sum rate as described in \cite{utschick2028lisaprecoding}.
The precoder is then applied to the full \gls{csi} to determine the resulting \gls{sinr}, which is used to compute the sum-rate as defined in \cite[Sec.~14]{cover2006inftheory}.
We report the sum-rates for each time step as the mean over all test samples and subcarriers.
Additional, we evaluate the alignment of dominant channel subspaces, by calculating the principal angles $\mathcal{E}_{\text{subspace},t,f,i}$ between the predicted and true right singular vectors for each dominant mode $i$ as discussed in \cite{bjorck1973angleerror}.
Based on these angles, we introduce a weighted subspace error
\begin{equation}
    \bar{\mathcal{E}}_{\text{subspace}t,f,}= \sum_{i=1}^{N_{\text{s}}} w_{t,f,i}\,\mathcal{E}_{\text{subspace},t,f,i}, \quad w_{t,f,i}=\frac{s_{t,f,i}}{\sum_{j=1}^{N_{\text{s}}} s_{t,f,j}},
\end{equation}
where $s_{t,f,i}$ denoting the $i$-th singular value.
This metric complements \gls{nmse} and sum-rates by capturing the subspace consistency.
Again, we report the mean over all test samples and subcarriers for each time step.
Lower values correspond to more accurate directions of the left singular vectors.

\subsection{Baseline Predictors}
As simple baselines for channel prediction, we consider \gls{zoh}, \gls{foh}, and \gls{lmmse}. 
\gls{zoh} considers the next time step $t$ to corresponding to the last observation, $\hat{\bm{z}}_{\text{p,ZOH},t} = \tilde{\bm{y}}_{\text{o},t-1}$.
\gls{foh} performs linear extrapolation from recent observations $\hat{\bm{z}}_{\text{p,FOH},t} = \tilde{\bm{y}}_{\text{o},t-1} + (\tilde{\bm{y}}_{\text{o},t-1} - \tilde{\bm{y}}_{\text{o},t-2})$.
Both, \gls{zoh} and \gls{foh} are computationally efficient but degrade with increasing mobility and prediction horizon \cite[Sec.~2]{oppenheim2010discrete}. 
\gls{lmmse} exploits second-order statistics under a Gaussian assumption.
The channel prediction is given as \cite[Sec.~9.1.6]{hossein2014nmselslmmsebook, turan2024gmmchannelprediction}
\begin{equation}
\begin{aligned}
    \hat{\bm{z}}_{\text{p,LMMSE}} &= \mathbb{E}[\bm{z}_{\text{p}} \mid \tilde{\bm{y}}_{\text{o}}]\\
    &=\bm{C}_{\text{S},\bm{z}_{\text{p}}\bm{z}_{\text{o}}} (\bm{C}_{\text{S},\bm{z}_{\text{o}}\bm{z}_{\text{o}}} + \sigma^2 \bm{I})^{-1} (\tilde{\bm{y}}_{\text{o}} - \bm{\mu}_{\bm{z}_{\text{o}}}) + \bm{\mu}_{\bm{z}_{\text{p}}},
\end{aligned}
\end{equation}
where $\bm{C}_{\text{S}} = \frac{1}{L} \sum_{\ell=1}^L \bm{z}^{(\ell)} \bm{z}^{(\ell),\operatorname{H}}$ denotes the sample covariance matrix of the data.
Lastly, we compare our approach to the  \gls{nn}-based method, which uses the \gls{gru} descibed in \cite{stenhammar2024nnchannelprediction}, which was shown to achieve robust performance in channel prediction.
The \gls{nn} contains two \gls{gru} layers with 128 hidden units, and is trained for 50 epochs using the Adam optimizer with a learning rate of $10^{-3}$, a complex-valued \gls{mse} loss, and a batch size of 256.
We compare the baselines with two \gls{gmm} variants, the first using a full covariance matrix and the second with a block-diagonal Toeplitz structure, under the independent-subcarrier processing discussed in Section~\ref{sec:block-diag-t}.
We evaluate the performance for training on full and partial \gls{csi}, to asses the impact of the partial format.

\subsection{Simulation Results}
Fig.~\ref{fig:tgn-nmse}, left side (a), shows the \gls{nmse} for $t_{\text{p}}=1$ and $\Delta t=20$ms, evaluated using only partial \gls{csi}.
\Gls{gmm}-based predictors outperform the classical baselines and shows a moderate improvement over the \gls{nn}-based approach, with partial feedback matching full feedback, demonstrating that the proposed models capture the essential channel structure even with reduced information.
Notably, the parameter-reduced block-diagonal Toeplitz (diag.-T.) \glspl{gmm} achieve performance comparable to \glspl{gmm} trained with a full covariance matrix (full-cov.), indicating the good quality of the approximation introduced in Section~\ref{sec:block-diag-t}.
Fig.~\ref{fig:tgn-nmse}, right side (b), shows the \gls{nmse} over varying $\Delta t$ at 30dB~\gls{snr}.
Again, \gls{gmm} predictors yield the lowest errors, with partial and full feedback having nearly identical performance and the parameter-reduced \glspl{gmm} exhibit no performance degradation.

\begin{figure*}[!t]
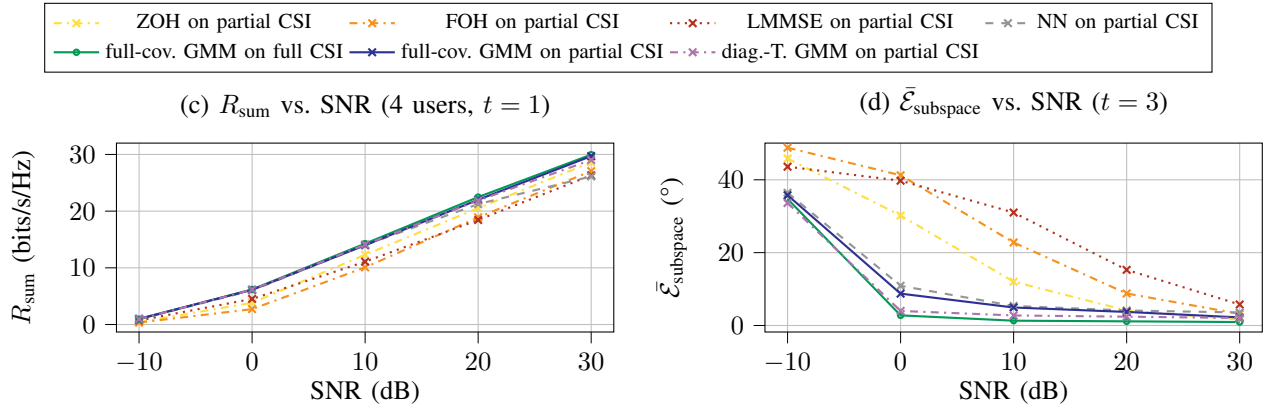
 
    \centering
    \begin{tikzpicture}
        \begin{groupplot}[
            group style={
                group size=2 by 1,
                horizontal sep=2cm,
            },
            width=0.45\textwidth,
            height=0.225\textwidth,
            legend to name=sharedLegendSmall,
            legend columns=4,
            legend style={fill opacity=0.8, draw opacity=1, text opacity=1, anchor=south, font=\footnotesize}
        ]
        \input{images/tgn_sum_rate_4_vs_snr_t1}
        \input{images/tgn_angle_error_vs_snr_t3}
        \end{groupplot}
    \end{tikzpicture}
    \caption{Evaluation of $R_{\text{sum}}$ for 4 users using LISA precoding and $\bar{\mathcal{E}}_{\text{subspace}}$ over \gls{snr}: $N_{\text{t,o}}=2$, $N_{\text{t,p}}=3$, $\Delta t = 20$ms, pred. step $t_{\text{p}}=\{1,3\}$ and $K=256$ for the \gls{gmm} variants.}
    \label{fig:tgn-mi-angles}
    \vspace{-0.6cm}
\end{figure*}

In Fig.~\ref{fig:tgn-mi-angles}, left side (c), we evaluate a four-user \gls{mu} scenario, using the partial predicted channels for \gls{lisa} precoding and assessing the resulting sum-rate.
\gls{gmm} variants consistently achieve the highest sum-rates.
Training on full or partial \gls{csi} yield to nearly identical results, demonstrating the reliability of \gls{gmm}-based predictors under partial feedback.
Once more, the block-diagonal Toeplitz \gls{gmm} shows to be an accurate approximation.
At high \glspl{snr}, the \gls{gmm} clearly outperforms the \gls{nn}, which performs worse than the classic baselines.

Considering the prediction step $t_{\text{p}}=3$, Fig.~\ref{fig:tgn-mi-angles}, right side (d), shows that \gls{gmm} predictors achieve the lowest weighted subspace error.
The \gls{gmm} trained on full \gls{csi} performs better, likely due to the evaluation procedure: the full \gls{csi} \gls{gmm} predicts the complete channel matrix, from which the corresponding singular vectors are subsequently extracted via \gls{svd}.
This post-processing step enforces the inherent orthogonality structure of the singular vectors. In contrast, the partial \gls{csi} \gls{gmm} directly predicts the quantities of interest without such structural enforcement.
Therefore, the comparison involves an additional reconstruction step for the full \gls{csi} approach, which contributes to its performance advantage.
The additional structure imposed by the block-diagonal Toeplitz (diag.-T.) variant of the \gls{gmm} allows to perform slightly better than the ones using full covariance matrices.
The gap to \gls{lmmse} grows substantially for $t_{\text{p}}=3$.
The advantage over other baselines, including the \gls{nn}, is particularly pronounced around 0dB~\gls{snr}, confirming that the probabilistic framework provided by the \gls{gmm} captures the temporal evolution of channel subspaces more accurately than classical linear predictors also for multi-step predictions.
The precise estimation of dominant directions explains why \gls{gmm} variants also achieve superior \gls{mu} sum-rate performance.

\begin{figure*}[!t] 
    \centering
    \ref{sharedLegend}
    \vspace{-0.4cm}
\end{figure*}

Across all settings, \gls{gmm}-based predictors outperform the baselines, especially for \gls{mu} setups we see an advantage compared to the \gls{nn}-based predictor.
The block-diagonal Toeplitz \glspl{gmm} using partial \gls{csi} matches the performance of a full \gls{csi} \gls{gmm}, showing that the reduced approach does not degrade performance.

\section{Conclusion and Outlook}
In this paper, we presented a prediction framework for forecasting future \gls{csi} in high-dimensional \gls{dl}-\gls{mu}-\gls{mimo}-\gls{ofdm} systems under feedback constraints.
By explicitly addressing partial feedback, the proposed approach remains applicable to realistic system settings, for example \gls{wifi} systems.
Furthermore, we introduced a parameter reduction technique that exploits structural properties of the covariance matrix, significantly lowering model complexity without degrading performance.
Simulation results demonstrated that the proposed method outperforms conventional channel predictors.

For future work, we will further explore structured covariance modeling and parameter reduction to efficiently handle large-scale \gls{mimo}-\gls{ofdm} systems, including approaches that explicitly enforce orthogonality in the predicted subspaces.
Evaluating alternative \gls{ml} models and extending simulations to real measurement data could further provide a more practical assessment of predictive performance.

\bibliographystyle{IEEEbib}
\bibliography{IEEEabrv,references}
\end{document}